\documentclass[prl,aps,floatfix,groupedaddress,showpacs,amsmath,twocolumn]{revtex4}  

\usepackage[dvips]{graphicx}
\usepackage{dcolumn}
\usepackage{bm}
\usepackage{epsfig}

\begin{document}

%%%%%%%%%%%%%%%%%%

\title{Diffusion Monte Carlo with lattice regularization}

\author{Michele Casula,$^{1,2}$ Claudia Filippi,$^3$ and Sandro Sorella$^{1,2}$ }

\affiliation{
$^1$ International School for Advanced Studies (SISSA) Via Beirut 2,4
  34014 Trieste, Italy \\
$^2$ INFM Democritos National Simulation Center, Trieste, Italy \\
$^3$ Universiteit Leiden, Instituut-Lorentz for Theoretical Physics,
  NL-2300 RA Leiden, The Netherlands
} 

\date{\today}

\begin{abstract}
We introduce an efficient lattice regularization scheme for 
quantum Monte Carlo calculations of realistic electronic systems.
The kinetic term is discretized by a finite difference 
Laplacian with two mesh sizes, $a$ and $a^\prime$, where $a^\prime/a$ is
an irrational number so that the electronic coordinates are not defined 
on a particular lattice but on the continuous configuration space.
The regularized Hamiltonian goes to the continuous 
limit for $a\to 0$ and provides several advantages.
In particular, it allows the inclusion of non-local potentials 
in a consistent variational scheme, substantially improving the accuracy 
upon previous non-variational approaches.  
\end{abstract}


\maketitle

In the last few decades, enormous progress has been made
in computing the physical properties of many-electron systems 
with numerical methods based on first principle quantum mechanics. 
In particular, quantum Monte Carlo (QMC) techniques~\cite{foulkesreview}
have proven very successful mainly because they allow the explicit 
inclusion of electronic correlation in a many-body wave function (WF). 
Moreover, projection QMC methods such as diffusion Monte Carlo 
(DMC) can improve upon a given guiding WF $\Psi_G$
by stochastically projecting a state $\Psi_{FN}$ much closer to 
the exact ground state (GS) of the Hamiltonian $H$.
$\Psi_{FN}$ is obtained within the fixed-node approximation (FNA),
which yields the lowest solution of the
Schr\"odinger equation with the same nodes as $\Psi_G$.
If $\Psi_G$ is appropriately optimized,
the method usually provides a good upper bound
$E_{FN}$ for the GS energy $E_{GS}$:
\begin{equation} \label{vmcfn}
E_{FN}={ \langle \Psi_{FN} | H | \Psi_{FN} \rangle \over 
\langle \Psi_{FN} | \Psi_{FN} \rangle }\,.
\end{equation}  
In general, in a QMC calculation, one accesses a mixed-average estimate
of the total energy:
\begin{equation} \label{ma}
E_{MA}=  {\langle \Psi_G | H | \Psi_{FN} \rangle \over
\langle \Psi_G | \Psi_{FN} \rangle }\,,
\end{equation}
which coincides with $E_{FN}$ since $\Psi_{FN}$ satisfies the equation
$H\Psi_{FN}=E_{FN}\Psi_{FN}$ within the nodal pockets of $\Psi_G$.

While the QMC methods can be extended to large systems containing 
many electrons, the computational 
effort increases dramatically with the atomic number $Z$. The most common 
way to overcome this difficulty is to replace the core electrons by 
pseudopotentials, an approximation which is usually rather good as
the core is chemically inert.
Since the pseudopotentials are in most cases non local, 
$H$ contains a non-local potential 
$V^P$, i.e.\ $\langle x |V^P| x^\prime \rangle  \ne 0$ even when 
$|x-x^\prime| \ne 0$, where $x$ denotes an all electron configuration 
with positions $\{\vec r_i\}$ and spins $\{\sigma_i\}$.
Consequently, the standard DMC approach cannot be applied and the so 
called ``locality approximation'' (LA) is
employed~\cite{hammond,hurley,christiansen,mitas}, which approximates
the non-local potential $V^P$ with a local one,
$V^{LA}(x) =  { \langle x |V^P | \Psi_G \rangle}  /
  {\langle x | \Psi_G \rangle }$.  
A major drawback of this approach is that 
$E_{FN}$ (Eq.~\ref{vmcfn}) is no longer available since it
does not coincide with 
$E_{MA}$ (Eq.~\ref{ma}) 
accessible in a DMC calculation: $\Psi_{FN}$ is now the best solution 
with the same nodes as $\Psi_G$ for the approximate Hamiltonian with 
potential $V^{LA}$ and not for the Hamiltonian $H$. Even though
$E_{MA}$ equals $E_{GS}$ if $\Psi_G$ is exact, it is otherwise
not bounded by $E_{GS}$ and no rigorous information about the quality 
of  $\Psi_{FN}$  is given by $E_{MA}$.
For instance, it is not possible to accertain the quality of the nodes of 
$\Psi_G$ since a lower $E_{MA}$ may correspond to a worse $\Psi_{FN}$,
namely with higher $E_{FN}$.
Empirically, the method appears to work~\cite{mitas}, but its drawback
has limited  the impact of this technique to 
a wider range of applications.

In this Letter, we present a lattice regularization of the many-electron 
Hamiltonian which removes the above difficulties when using non-local 
potentials within the FNA. We demonstrate the utility of our lattice 
regularized DMC approach in cases where the LA yields inaccurate results.

{\it Regularization of the Hamiltonian}. 
We consider the Hamiltonian in atomic units:
\begin{equation} \label{hamuniverse}
H= -\frac{1}{2} \sum_i \Delta_i + V(x)\,,
\end{equation}
and, for the moment, assume a local potential,
$V(x)=  \sum_j -Z_j /|\vec r_i- \vec R_j| 
+ \sum_{i<j} 1/| \vec r_i - \vec r_j |$, where $\vec R$ and 
$\vec r$ indicate the ionic and electron positions, respectively.

The Laplacian is approximated by a finite difference form   
\begin{equation} \label{defnablaa} 
\Delta_i \approx\Delta^a_i=\Delta^{a,p}_i + \Delta^{a',1-p}_{i} +O(a^2)\,,
\end{equation} 
where $\Delta^{a,p}$ is an appropriate Hermitian lattice operator 
defined by a mesh size $a$, a constant $\eta$, and an arbitrary   
function $p(\vec r_i)$:
\begin{eqnarray*} \label{laplace}
&\Delta^{a,p}_i &  f(x_i,y_i,z_i)=\eta/a^2 \left[ p(x_i + a/2)
    (f(x_i+a)-f(x_i))  \right.  \\ 
&+&\left.  p(x_i-a/2)  (f(x_i-a)-f(x_i))\right]   +   
x_i\leftrightarrow y_i \leftrightarrow z_i.  	 
\end{eqnarray*}
For $p=\eta=1$, $\Delta^{a,p}_i$ coincides with the usual 
discretized form of the Laplacian on a lattice with mesh size $a$. 
By combining two such operators in Eq.~\ref{defnablaa} 
with mesh sizes $a$ and $a'$, and $a^\prime/a$ an irrational number,
the electron coordinates $\{ \vec r_i \}$ are still defined in the 
continuous space because the two meshes are incommensurate and the 
diffusion process described by this discretized Laplacian 
can cover all the configuration space.
Moreover, in Eq.~\ref{defnablaa}, the function $p$ can be arbitrarily 
spatial dependent as long as $0\le p \le 1$ to ensure that $p$ and $1-p$ 
are both positive.
In order to optimize the efficiency of the diffusion process,
we have carefully chosen the constant $a^\prime/a$ and the function $p$
as:
\begin{subequations}
\begin{equation}
p(\vec r)= 1/(1+ Z^2/ 4|\vec r - \vec R|^2 )\,, 
\end{equation}
\begin{equation}
a'/a=\sqrt{Z^2/4+1}\,,
\end{equation}
\end{subequations}
where $\vec R$ and $Z$ are the position and the atomic 
number of the nucleus closest to the electron in $\vec r$.
Therefore, if the electron is very close to a  nucleus, the smaller
mesh $a$ is used, while, in the valence region where $|\vec r -\vec R|>>1/Z$, 
the electron can make steps of larger amplitude $a'$, 
thus reducing the QMC correlation time. 

If $\eta=1 +O(a^2)$, the resulting finite difference 
Laplacian $\Delta^a_i$ coincides with the continuous $\Delta$ 
up to $O(a^2)$.
To further improve the accuracy of the approximation and 
work with reasonably large value of $a$,  we regularize
also the potential $V \to V^a$ so that our final regularized 
Hamiltonian, $H^a= -{ 1/2} \sum_i \Delta^a_i + V^a $, fulfills 
the following three conditions: 
i) $H^a \to H$ for $a \to 0$;
ii) for the  chosen  guiding WF $\Psi_{G}$,
for any $a$ and any $x$,
the local energy $e_L (x,[\Psi_{G}]) = { \langle x | H | \Psi_{G}
\rangle} /{ \langle x | \Psi_{G} \rangle } $ of the continuous 
Hamiltonian $H$ is equal to $e^a_L(x,{[\Psi_{G}]})$ corresponding 
to the Hamiltonian $H^a$;
iii) the discretized kinetic energy is equal to 
the continuous one calculated on the state  $\Psi_{G}$, i.e.\
$\langle -1/2 \sum_i  \Delta^a_i \rangle_{\Psi^2_G}  = 
\langle -1/2 \sum_i \Delta_i \rangle_{\Psi^2_G}.$
While the condition (iii) determines the constant $\eta$, 
the condition (ii) leads to the following solution for $V^a$: 
\begin{equation} \label{secondreg}
 V^a(x) = V(x)+  \frac{1}{2 } \left[ \frac{\sum_i ( \Delta^a_i - \Delta_i)
     \Psi_{G}}  
{\Psi_{G}}\right](x).
\end{equation}
Notice that the condition (ii) yields another important property
for $H^a$:
if $\Psi_{G}$ is an eigenstate of $H$, it is also  
an eigenstate of $H^a$ for any $a$, as  can be derived 
by simple inspection using that $\Delta^a$ is Hermitian.  
Thus, by improving $\Psi_{G}$, a better $a\to 0$ convergence 
is also expected.  

{\it Non-local pseudopotential}.
The above regularization also applies when we include in the potential
$V$ a non-local potential of the form $V^P=\sum_i v^P(\vec r_i)$ 
where
\begin{equation} \label{pseudo}
v^{P}  (\vec r_i) =\sum_{j} \bigg[
v_{j}(|\vec r_i -\vec R_{j}|) +  \sum_{l \le l_{max}^j}  v_j^{l} (|\vec r_i-
\vec R_{j}| ) 
P_l \bigg]\,.
\end{equation}
The sum over $j$ is here restricted to the pseudoatoms,
$P_l$ is the $l$-angular-momentum projection centered on the 
pseudoatom at $\vec R_{j}$, and $l_{max}^j$ the corresponding maximum 
angular momentum considered. The functions $v_{j}$ and $v_{j}^l$ are radial, 
and $v_{j}^l$ vanishes outside a core radius $r_c^{j}$. 
  
In a QMC calculation with pseudopotentials~\cite{fahy,mitas}, the 
angular integration to evaluate the projection $P_l$ is performed 
numerically on a regular polyhedron defined by $N_V$ vertices and
by employing a quadrature rule for the integration on 
a sphere.
Therefore, the non-local pseudopotential $v^P$ acts on a 
configuration $x$ by  means of 
a {\em finite} number of matrix elements equal to $N_V N_{core}$,
where $N_{core}$ is the number of electrons within the core radius
of a pseudoatom.
In the following, we will assume that, in the continuous Hamiltonian $H$,
the pseudopotential is by definition discretized with 
a finite number $N_V$ of vertices. Then, the regularization $H \to H^a$ 
follows by simple substitution of $V(x) \to V(x)+V^P$ in Eq.~\ref{secondreg}.

{\it Lattice regularized diffusion Monte Carlo} (LRDMC).
Although $H^a$ is an Hamiltonian defined 
on a continuous space, all techniques valid on a lattice can be 
straightforwardly applied here since $H^a$ acts on a configuration 
exactly as a lattice Hamiltonian, namely: 
\begin{equation} \label{property}
%H^a |x \rangle = \sum_{x^\prime}  H_{x^\prime,x}^a |x^\prime\rangle\,,
\langle x | H^a | \Psi_G \rangle = \sum_{x^\prime}  H_{x,x^\prime}^a \langle
x^\prime | \Psi_G \rangle\,,
\end{equation}
where, for a given $x$, the number of matrix elements $H_{x,x^\prime}^a$
are \emph{finite} and can be computed even in the presence 
of non-local pseudopotentials. In particular, we can resort to the same scheme
used in the efficient lattice Green function Monte Carlo
algorithm~\cite{ceperley,calandra,capriotti}.
The resulting algorithm, valid for a continuous regularized
Hamiltonian, is the LRDMC.
The corresponding Green function matrix elements are
$G_{x^\prime,x}=\Psi_G(x^\prime)(\Lambda\delta_{x^\prime,x}- 
H_{x^\prime,x}^a)/\Psi_{G} (x)$, and, provided they are all non-negative,
the positive distribution $\Psi_G(x) \Psi_{GS} (x)$ is statistically sampled.
Note that, since the spectrum of $H^a$ is not bounded from above, we need to 
take the limit $\Lambda\to\infty$, which can be handled with no loss of
efficiency as described in Ref.~\onlinecite{capriotti}.  
The LRDMC algorithm is very simple:
given a walker with configuration $x$ and weight $w$, a 
new configuration  $x^\prime \ne x$ is obtained with probability
$p_{x^\prime,x}= G_{x^\prime,x} / N_x $, 
where $N_x= \sum_{x^\prime (\ne x)} G_{x^\prime,x} $ is the normalization.
The walker weight is then changed by a factor 
$w \to w\, \exp(- \tau_x e_L (x,[\Psi_G]) ) $ 
where $\tau_x= -log(r) /N_x$ is a diffusion time  determined by  a
single random number $0< r \le 1$.
Then, the usual branching scheme is used to control the 
fluctuations of the walker weights.

Within this approach, the time-step error of the standard DMC method is 
not present, but is replaced by the error of the Laplacian discretization.
By performing LRDMC simulations of $H^a$ for different $a$'s, 
we can approach the GS properties of a continuous system 
in the limit $a\to 0$.
The method becomes less efficient as $a\to 0$ since
the computational time increases like $1/a^2$, similarly to the DMC  
efficiency 
which is limited by $1/\Delta \tau$, where $\Delta \tau$ is the imaginary 
time step for the diffusion process.

\begin{figure}[tb]
\epsfxsize=85mm
\centerline{
\epsffile{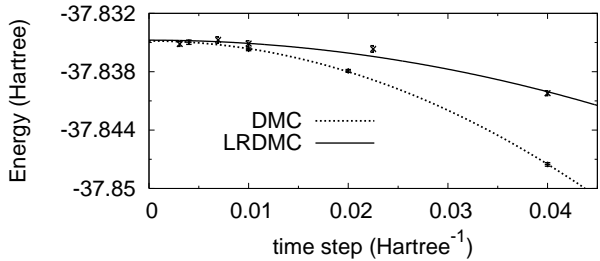}}
\caption{\label{allel} 
FN energies for the all-electron carbon atom computed within DMC and LRDMC. 
The lattice spacing $a$ has been mapped 
to the time-step $\tau$ using the relation $a=\sqrt{\tau}$.
}
\vspace{-0.2cm}
\end{figure}

Since the Green function $G_{x^\prime,x}$ can be made strictly positive 
only for bosons, we have to introduce here the analogous 
of the FNA on a lattice~\cite{ceperley,calandra,capriotti} by modifying
few of the matrix elements of the regularized Hamiltonian $H^a$.
For each configuration $x$, the matrix elements $H_{x^\prime,x}^a$ which
yield $G_{x^\prime,x} < 0$ are set to zero and included in the so called 
sign-flip term, ${\cal V}_{sf} (x)=\sum_{x^\prime \ne x },
\Psi_G(x^\prime) H_{x^\prime,x}^a/\Psi_G (x) > 0$, which is then
added to the diagonal element $H_{x,x}^a$~\cite{ceperley}. The resulting 
effective 
Hamiltonian $H^{\textrm{eff}}$ has the same local energy as $H^a$,
it is non frustrated, and its GS WF
has the same signs as the guiding function $\Psi_G(x)$. 
The GS energy of $H^{\textrm{eff}}$ 
can be efficiently computed with the mixed average $E_{MA}$
and, for a local Hamiltonian $H$, we recover the 
standard DMC result $E_{MA}=E_{FN}$ in the limit $a\to 0$ 
as shown in Fig.~\ref{allel}.
Moreover, when $H$ contains non-local pseudopotentials,
$E_{MA}$ for the GS of $H^{\textrm{eff}}$ is certainly lower than
the expectation value $E_G$ of the Hamiltonian $H^a$ on $\Psi_G$, and 
represents an upper bound of the variational energy $E_{FN}$ with the 
Hamiltonian $H^a$ in Eq.~\ref{vmcfn}, as was shown for the lattice FN 
approach~\cite{ceperley}.
This important upper bound property has been so far only obtained 
within the lattice regularization proposed here, whereas,
in the DMC approach, it is not true in general that 
$E_G \ge E_{MA} \ge E_{FN}$. 

%%%%%%%%%%%%%%%%%%%%%%%%%%%%%%%%%%%%%%%%%%%%%%%%%%%%%%%%%%%%%%%%%%%%%%%%%
%%%%%%%%%%%%%%%%%%%%%%%%%%%%%%%%%%%%%%%%%%%%%%%%%%%%%%%%%%%%%%%%%%%%%%%%%%

{\it Computation of $E_{FN}$}. Another advantage of 
our lattice regularization is that we can also compute directly and accurately 
$E_{FN}$ for the Hamiltonian $H$ by introducing a more general effective 
Hamiltonian characterized by the parameters $\gamma \ge 0$ 
and $ 0\le  \alpha <1$~\cite{sorellaeff}:
\begin{eqnarray}
H^{\textrm{eff}}_{x,x} & = & H^a_{x,x} + (1+\gamma) {\cal V}_{sf}(x) 
+\alpha (1+\gamma)  {\cal V}_{sf}^P(x) \\
H^{\textrm{eff}}_{x^\prime,x} & = & - \gamma H^a_{x^\prime,x} 
\textrm{~~~~~~~~~~~~~~~~~if $\Psi_G(x^\prime) H^a_{x^\prime,x} / \Psi_G(x) >
  0$}\,\nonumber \\ 
H^{\textrm{eff}}_{x^\prime,x} & = & (1-\alpha (1+\gamma) ) H^a_{x^\prime,x} 
\textrm{~~if $\Psi_G(x^\prime) V^P_{x^\prime,x} / \Psi_G(x) < 0$} \nonumber  \\
H^{\textrm{eff}}_{x^\prime,x} & = & H^a_{x^\prime,x} 
\textrm{~~~~~~~~~~~~~~~~~~~~~otherwise}, \nonumber
\end{eqnarray}
where $x^\prime \ne x$ and ${\cal V}_{sf}^P(x) =\sum_{x^\prime \ne x}  
 \Psi_G(x^\prime) V^P_{x^\prime,x} / \Psi_G(x)<0$.
This Hamiltonian also satisfies $G_{x^\prime,x} > 0$ and reduces
to the standard lattice FN one for $\alpha=\gamma=0$. 

The parameter $\gamma$ allows one to evaluate $E_{FN}$ in the
presence of the non-local contribution of $V^P$ as
\begin{equation}
E_{FN}=E_{MA}(\gamma)-(\gamma+1) 
\frac{d E_{MA}(\gamma)} {d \gamma}\,,
\end{equation}
where we used that 
$H=H^\textrm{eff} - (\gamma+1) \partial_\gamma H^\textrm{eff}$
and applied the Hellmann-Feynman theorem to the last term.
To estimate $E_{FN}$, we compute the derivative with respect to $\gamma$
in an approximate but variational way, the best variational estimate 
being for $\gamma=0$~\cite{sorellaeff}:
\begin{equation}\label{2ma}
E_{FN} \le E_{MA}(0) - {
\left[E_{MA}(\gamma)-E_{MA}(0)\right]}/{\gamma},
\end{equation}
where the equality sign holds in the limit of small $\gamma$. 
Notice that, for $\alpha=1$ and $\gamma=0$, the LA is recovered, 
i.e.\ all the 
off-diagonal matrix elements of $V^P$ are included in the diagonal 
ones of $H^{\rm eff}$. Even though it is not garanteed that
$E_{MA}$ is variational for $\alpha>0$,
with the present scheme we can evaluate $E_{FN}$
and improve upon the locality approximation by optimizing
$\alpha$. Indeed, 
the variational energy $E_{FN}$ corresponding to a given value 
of $\alpha$ can be estimated efficiently using correlated sampling to evaluate 
$E_{MA}$ for $\gamma=0$ and $0< \gamma \le 1/\alpha -1$ as in Eq.~\ref{2ma}. 
Remarkably, this approach allows us to work close to the locality 
approximation, even with $\alpha =0.95$ and $\gamma=0.05$, 
with statistical errors for $E_{FN}$ similar to the ones 
for $E_{MA}$. 

{\it Results and perspectives}.
We have first tested the performance of the LRDMC approach on the silicon 
pseudoatom using three Hartree-Fock pseudopotentials which differ in the
construction, functional form and core radius.
For each pseudopotential, we employ three WF's with the 
same determinantal component and, consequently, the same nodes, but with 
different Jastrow factors. We use no Jastrow factor, a two-body, and a 
sophisticated three-body Jastrow factor~\cite{jastrowfilippi}. 

\begin{figure}[tb]
\epsfxsize=85mm
\centerline{
\epsffile{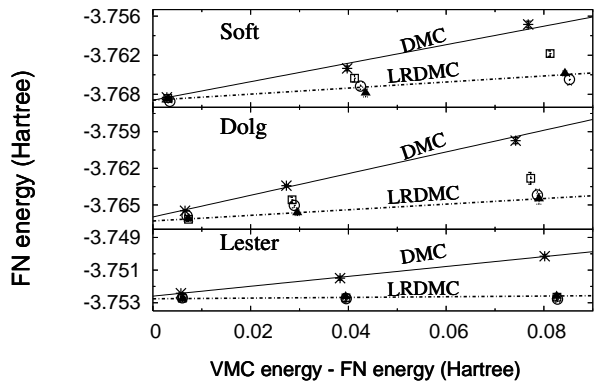}}
\caption{\label{sipseudo} 
FN energies of the silicon pseudoatom computed within DMC and LRDMC. 
For different pseudopotentials (Soft~\cite{shirley}, Dolg's~\cite{dolgpseudo} 
and Lester's~\cite{lesterpseudo}), 
we use as guiding WF's a Hartree-Fock determinant 
with no Jastrow, a two-body and a three-body Jastrow factor.
A more accurate guiding WF corresponds to a smaller difference between the 
variational Monte Carlo (VMC) and the FN energies. 
The LRDMC energies are computed for $\alpha=0.9$ (filled triangles), 
$\alpha=0.5$ (open circles) and $\alpha=0$ (open squares). 
The linear fits for the DMC and the LRDMC ($\alpha=0.9$) data are shown.
}
\vspace{-0.2cm}
\end{figure}

As shown in Fig.~\ref{sipseudo}, the energy estimate $E_{MA}$ computed 
within DMC changes significantly with the guiding WF $\Psi_G$, 
and differs from the variational expectation value, $E_{FN}$, 
which we compute with the LRDMC scheme for $\alpha=0$, $0.5$ and $0.9$.
For all cases, the statistical uncertainty does not allow us to discriminate 
between the LRDMC energies obtained for $\alpha=0.5$ and $\alpha=0.9$, and a 
shallow minimum seems to lie between these two values. The optimal $E_{FN}$ 
is almost free of the localization error and the weakest
dependence on $\Psi_G$ is obtained for Lester's 
pseudopotential~\cite{lesterpseudo} which has the smallest core radius in 
the non-local component. 
Interestingly, since $E_{FN}$ for $\alpha \simeq 1$ is very close to 
the minimum, the LA seems to yield good WF's. However, the value 
of $\alpha$ should be optimized for each case since there is no general 
reason why $\alpha=1$ should give the best energy. In principle, in the case
of systems with lower symmetry, the value of
$\alpha$ should be closer to $0$ due to the non conservation of 
the total angular momentum.

%%%%%%%%%%%%%%%%%%%%%%%%%%%%%%%%%%%%%%%%%%%%%%%%%%%%%%%%%%%%%%%%%%%%%%%%%%%%%
%%%%%%%%%%%%%%%%%%%%%%%%%%%%%%%%%%%%%%%%%%%%%%%%%%%%%%%%%%%%%%%%%%%%%%%%%%%%%

A stringent test case for our LRDMC algorithm is the scandium atom:
the LA for transition metals yields large errors in the DMC total 
energies, and performs the worst for the scandium atom~\cite{dolgmetal}.
As before, we keep the determinantal part of the WF fixed,
and employ a 2-body~\cite{jastrowcusp} and a 3-body~\cite{casula} Jastrow factor.
The determinantal component is an antisymmetrized geminal function expanded 
over a ($5s5p5d$) Gaussian-type basis in order to cure near degeneracy effects, 
and optimized in the presence of the 2-body Jastrow factor.
We employ Dolg's pseudopotential~\cite{dolgmetal} and compute the 
$4s^2 3d^n \rightarrow 4s^1 3d^{n+1}$ excitation energy 
wich is reported in Table~\ref{sc}. 
It is apparent that the LA does not only affect the DMC total energies but also
the DMC energy differences: the DMC excitation energy computed with the 2-body 
Jastrow factor differs from the experimental value by more than three standard 
deviations. On the other hand, the LRDMC FN results, obtained with $\alpha=0.5$,
are much less sensitive to $\Psi_G$, and are compatible with the experiment 
even when a simple 2-body Jastrow factor is employed.
The LRDMC mixed-average excitations are also closer to the experimental value
than  
the LA ones, probably because the variational treatment of non-local
pseudopotentials gives a better cancellation of errors. Remarkably, we found
that a LRDMC simulation with off-diagonal pseudopotentials is computationally
much more stable and 
efficient than the standard DMC approach since the negative divergences
coming from the pseudopotentials close to the nodes are converted to finite
hopping terms in the LRDMC scheme.

\begin{table}[htb]
\begin{ruledtabular}
\begin{tabular}{l|c|c|c|c|c|c}
         &\multicolumn{1}{c}{}
         &\multicolumn{1}{c}{}
         &\multicolumn{1}{c}{}
         &\multicolumn{2}{c}{LRDMC} &\\
\cline{5-6}
         &\multicolumn{1}{c}{$\alpha$}
         &\multicolumn{1}{c}{VMC}
         &\multicolumn{1}{c}{DMC}
         & \multicolumn{1}{c}{$E_{MA}(\gamma=0)$} 
         & \multicolumn{1}{c}{$E_{FN}$} 
         & \multicolumn{1}{c}{exp} \\
\hline
2-body         & 0.0 & 1.099(30) & 1.369(19) & 1.407(16) & 1.416(43) & 1.43\\
2-body         & 0.5 & 1.099(30) & 1.369(19) & 1.399(16) & 1.452(34) &    \\
3-body         & 0.5 & 1.303(29) & 1.445(38) & 1.448(16) & 1.492(41) &    \\
\end{tabular}
\end{ruledtabular}
\caption{Comparison of 
 $4s^2 3d^n \rightarrow 4s^1 3d^{n+1}$ excitation energy (eV) for the
  scandium atom.}
\label{sc}  
\vspace{-0.2cm}
\end{table}

We have presented an efficient lattice regularization scheme 
for QMC calculations on realistic electronic systems.  The main 
advantage of the  LRDMC approach is the possibility to work with
non-local potentials within a fully consistent and variational scheme
which is much more accurate than the standard DMC method. 
Moreover, this projection method allows one
to deal with several length scales through the use of multiple
lattice spaces, with great reduction of autocorrelation times 
for heavy atoms or complex systems.
We believe that this framework can have a wide spread of important
applications ranging from nuclear physics~\cite{fantoni} to
the chemistry of transition metal compounds.

This work was partially supported by INFM, by MIUR (COFIN 2003),
and by Stichting voor Fundamenteel Onderzoek der Materie (FOM).
We acknowledge G. Bachelet and S. De Gironcoli for useful discussions.


\begin{thebibliography}{99}
\bibitem{jastrowcusp} The pseudopotentials of Ref.\ [15] diverge at the origin 
as -$Z_{\rm eff}/r$, so our two-body Jastrow factor has the form 
$\exp(\frac{ar}{1+br})$ to satisfy the cusp conditions for the
nucleus ($a=-Z_{\rm eff}$) and the antiparallel electrons ($a=1/2$).
\end{thebibliography}
\end{document}